\documentclass[manuscript,twocolumn]{aastex701}

\newcommand{\eps}{erg\,s$^{-1}$}

\newcommand{\M}{$M_{\odot}$~}

\usepackage{amsmath}	% Advanced maths commands
\usepackage{amssymb}	% Extra maths symbols
\usepackage{orcidlink}

\begin{document}

%\title{Changing-State Active Galactic Nuclei: III. Fraction of Coronal X-ray Emission and Eddington Ratio}
%\title{Changing-state Active Galactic Nuclei reveal the connection between X-ray and optical emission}
%\title{Changing-state Active Galactic Nuclei: III. The Connection between X-ray and Optical Emission in Accreting Supermassive Black Holes}
%\title{Extreme variability reveals the connection between X-ray and optical/UV emission in accreting supermassive black holes}
\title{Extreme Variability Reveals How the Eddington Ratio Regulates Coronal Power in Active Galactic Nuclei}

\author[orcid=0000-0001-7500-5752,sname='Jana']{Arghajit Jana}
%\altaffiliation{SRM University-AP}
\affiliation{Department of Physics, SRM University-AP, Amaravati, Andhra Pradesh, 522240, India}
\email[show]{arghajit.j@srmap.edu.in}

\author[0000-0001-5231-2645]{Claudio Ricci}
%\altaffiliation{University of Geneva}
\affiliation{Department of Astronomy, University of Geneva, ch. d'Ecogia 16, 1290 Versoix, Switzerland}
\affiliation{Instituto de Estudios Astrof\'isicos, Facultad de Ingenier\'ia y Ciencias, Universidad Diego Portales, Av. Ej\'ercito Libertador 441, Santiago, Chile}
\affiliation{Kavli Institute for Astronomy and Astrophysics, Peking University, Beijing 100871, China}
\email[show]{claudio.ricci@unige.ch}

\author[0000-0002-8686-8737]{Franz E. Bauer}
%\altaffiliation{Universidad de Tarapac\'a}
\affiliation{Instituto de Alta Investigaci\'on, Universidad de Tarapac\'a, Casilla 7D, Arica 1010000, Chile}
\email{franz.e.bauer@gmail.com}

\author[0009-0007-9018-1077]{Kriti Kamal Gupta}
\affiliation{STAR Institute, Li\`ege Universit\'e, Quartier Agora, All\'ee du Six Ao\^ut 19c, B-4000 Li\`ege, Belgium}
\affiliation{Sterrenkundig Observatorium, Universiteit Gent, Krijgslaan 281 S9, B-9000 Gent, Belgium}
\affiliation{Leibniz-Institut für Astrophysik Potsdam (AIP), An der Sternwarte 16, D-14482 Potsdam, Germany}
\email{kriti.gupta@mail.udp.cl}

\author[0000-0002-3683-7297]{Benny Trakhtenbrot}
\affiliation{School of Physics and Astronomy, Tel Aviv University, Tel Aviv 69978, Israel}
\email{bennyt@tauex.tau.ac.il}

\author[0000-0003-3450-6483]{Alessia Tortosa}
%\affiliation{INAF---Osservatorio Astronomico di Roma, Via di Frascati 33, I-00078 Monte Porzio Catone, Italy}
\affiliation{INAF - Osservatorio di Astrofisica e Scienza dello Spazio di Bologna, via Piero Gobetti, 93/3, I-40129 Bologna, Italy}
\email{alessia.tortosa@inaf.it}
%\email{alessiatortosa@gmail.com}

\author[0000-0001-8496-4162]{Ruancun Li}
\affiliation{Max-Planck-Institut f\"ur extraterrestrische Physik, Gie\ss enbachstra\ss e 1, 85748 Garching, Germany}
\email{liruancun@gmail.com}

\author[0000-0001-6947-5846]{Luis C. Ho}
\affiliation{Kavli Institute for Astronomy and Astrophysics, Peking University, Beijing 100871, China}
\affiliation{Department of Astronomy, School of Physics, Peking University, Beijing 100871, China}
\email{lho.pku@gmail.com}

\author[0000-0002-7962-5446]{Richard Mushotzky}
\affiliation{Department of Astronomy and Joint Space-Science Institute, University of Maryland, College Park, MD 20742, USA}
\email{rmushotz@umd.edu}

\author[0000-0002-5617-3117]{Hsiang-Kuang Chang}
\affiliation{Institute of Astronomy, National Tsing Hua University, Hsinchu 300044, Taiwan}
\email{hkchang@mx.nthu.edu.tw}

\author[0000-0002-8604-1158]{Yaherlyn Diaz}
%\affiliation{Instituto de Estudios Astrof\'isicos, Facultad de Ingenier\'ia y Ciencias, Universidad Diego Portales, Av. Ej\'ercito Libertador 441, Santiago, Chile}
\affiliation{Departamento de Física, Universidad Técnica Federico Santa María, Vicuña Mackenna 3939, San Joaquín, Santiago, Chile}
\email{yaherlyn.diaz@mail.udp.cl}

\author[0009-0002-4945-5121]{Georgios Dimopoulos}
\affiliation{NASA Goddard Space Flight Center, Code 662, Greenbelt, MD 20771, USA}
\affiliation{Instituto de Estudios Astrof\'isicos, Facultad de Ingenier\'ia y Ciencias, Universidad Diego Portales, Av. Ej\'ercito Libertador 441, Santiago, Chile}
\email{georgios.dimopoulos@mail.udp.cl}

\author[0009-0008-8860-0372]{Krist\'ina Kallov\'a}
\affiliation{Instituto de Estudios Astrof\'isicos, Facultad de Ingenier\'ia y Ciencias, Universidad Diego Portales, Av. Ej\'ercito Libertador 441, Santiago, Chile}
\email{kristina.kallova@mail.udp.cl}

\author[0000-0001-8433-550X]{Michael J. Koss}
\affiliation{Eureka Scientific, 2452 Delmer Street, Suite 100, Oakland, CA 94602-3017, USA}
\affiliation{Space Science Institute, 4750 Walnut Street, Suite 205, Boulder, CO 80301, USA}
\email{mike.koss@eurekasci.com}

\author[0000-0002-8108-9179]{St\'ephane Paltani}
\affiliation{Department of Astronomy, University of Geneva, ch. d'Ecogia 16, 1290 Versoix, Switzerland}
\email{stephane.paltani@unige.ch}

\author[0000-0001-8433-550X]{Matthew J. Temple}
\affiliation{Centre for Extragalactic Astronomy, Department of Physics, Durham University, South Road, Durham DH1 3LE, UK}
\email{matthew.j.temple@durham.ac.uk}

\begin{abstract}
{The bolometric luminosity ($L_{\rm bol}$) of active galactic nuclei (AGNs) is a key tracer of accretion physics, but its direct determination is often hindered by limited spectral coverage and contamination of the host galaxy. Bolometric corrections ($\kappa_{\lambda} = L_{\rm bol}/L_{\lambda}$) offer a practical means of estimating $L_{\rm bol}$, with the X-ray bolometric correction ($\kappa_{\rm 2-10}$) being crucial for exploring the coupling between the accretion disk and the X-ray corona. Here we present multi-epoch, multi-wavelength observations of five highly variable, changing-state AGNs that span more than three orders of magnitude in Eddington ratio ($-3.6\lesssim \log \lambda_{\rm Edd} \lesssim -0.5$). This unique data set reveals a remarkably tight relation between $\kappa_{\rm 2-10}$ and $\lambda_{\rm Edd}$, with an intrinsic scatter of only $\sim0.05$ dex. We find that while the sources show bolometric corrections following different tracks in luminosity space that depend on black hole mass, they all display the same $\kappa_{\rm 2-10}$–$\lambda_{\rm Edd}$ trend. This shows unambiguously that $\lambda_{\rm Edd}$ is the primary driver of X-ray bolometric corrections, and points to a tight underlying trend that can be used to obtain reliable estimates of bolometric output from X-ray luminosities. Our results highlight how time-domain, multi-wavelength observations of variable AGN offer unique insights into the accretion flow structure and its radiative output.}

\end{abstract}


\section{Introduction}
\label{sec:intro}

Active Galactic Nuclei (AGNs) are powered by the accretion of matter onto supermassive black holes (SMBHs) at the centers of galaxies (e.g., \citealp{Ress1984}). The gravitational potential energy of the infalling matter is converted into electromagnetic radiation that spans the entire spectrum, from radio to gamma rays. Matter accretes onto the SMBH through a spiraling accretion disk, which is thought to predominantly emit in the UV/optical wavebands \citep{SS73,NT73}. A fraction of these UV/optical seed photons are then upscattered via inverse Comptonization in a cloud of hot electrons, known as the X-ray corona, located close to the SMBH, producing X-ray continuum radiation \citep{ST80,ST85,HM1991,CT95,Fabian2015}.
 
Accurately measuring the total energy output, or bolometric luminosity ($L_{\rm bol}$), is crucial to understanding the physics of accretion, black hole growth, and AGN feedback \citep{Marconi2004,Kormendy2013}. However, direct determination of $L_{\rm bol}$ is observationally challenging due to incomplete spectral coverage, non-simultaneous observation in different bands, contamination from host galaxy starlight, and obscuration \citep[e.g.,][]{Elvis1994,Lusso2012,Netzer2013}. To overcome these challenges, bolometric correction factors are widely used. These are scaling factors that convert monochromatic or band-limited luminosities ($L_{\rm \lambda}$) to $L_{\rm bol}$ \citep[e.g.,][]{Richards2006,Runnoe2012}. The bolometric correction in a given waveband is defined as $\kappa_{\rm \lambda} = L_{\rm bol} / L_{\rm \lambda}$ and provides information on the fraction of total energy emitted in a particular energy band.

Bolometric corrections vary depending on the wavelength and accretion properties of the AGN. In the optical band, bolometric corrections tend to be relatively stable across AGN populations, showing little dependence on black hole mass ($M_{\rm BH}$), $L_{\rm bol}$, or Eddington ratio ($\lambda_{\rm Edd}=L_{\rm bol}/L_{\rm Edd}$; $L_{\rm Edd}=1.5\times10^{38}$ erg\,s$^{-1} (M_{\rm BH}/M_{\odot})$ is Eddington luminosity;  \citealp{Netzer2019,Runnoe2012,Duras2020}).
In contrast, bolometric corrections in the X-ray band show strong dependence on some fundamental AGN properties. In particular, the X-ray bolometric correction in the 2--10\,keV band ($\kappa_{\rm 2-10}$) has been observed to rapidly increase with both $L_{\rm bol}$ and $\lambda_{\rm Edd}$ \citep{Vasudevan2009,Lusso2012,Duras2020,Gupta2025}, albeit with large scatter. This trend shows that the relative contribution of the X-ray-emitting corona decreases as the accretion rate increases, with more energy being emitted in the UV-optical regime.
This trend indicates that the connection between the X-ray corona and the accretion flow evolves with the Eddington ratio, likely driven by enhanced cooling, regulated by the pair thermostat, or due to changes in the overall structure of the accretion flow \citep[e.g.,][]{Shemmer2008,Done2012,Ricci2018}.
The X-ray bolometric correction is particularly important, since X-ray emission is widely used as a proxy of the SMBH accretion rate. Understanding the physical drivers of $\kappa_{\rm 2-10}$ remains challenging as both $L_{\rm bol}$ and $\lambda_{\rm Edd}$ are derived from the 2--10\,keV X-ray continuum luminosity ($L_{\rm PC}^{\rm 2-10}$), making it difficult to unravel their individual effects.

Changing-state AGNs (CSAGNs) are ideal laboratories to study the disk-corona coupling and bolometric corrections because they show large amplitude flux variability. CSAGNs exhibit dramatic spectral transitions, switching between Type\,1 to Type\,2 optical classifications, over months to years (e.g., \citealp{LaMassa2015,MacLeod2016,Ricci2020}; see \citealp{Ricci2023Nat} for a recent review). These changes are typically associated with rapid changes in the accretion rate (e.g., \citealp{Ricci2023Nat,Temple2023}). In CSAGNs, the accretion rate can change by large amplitude on a timescale of months to years \citep[e.g.,][]{Stern2018,Noda2018,Ricci2023Nat}, providing a unique opportunity to investigate how $\kappa_{\rm 2-10}$ is affected by changes in luminosity and $\lambda_{\rm Edd}$. Multi-epoch, multi-wavelength observations of these systems can therefore reveal the physical driver of the fraction of the energy output in the X-ray corona.

In this work, we study $\kappa_{\rm 2-10}$ simultaneous optical/UV and X-ray spectra of five CSAGN, observed over 1000 epochs, exploiting their extreme variability to understand whether $L_{\rm bol}$, $\lambda_{\rm Edd}$ or $M_{\rm BH}$ control $\kappa_{\rm 2-10}$.

\section{Sample and data}
\label{sec:data}
Here we focus on a subset of five extremely variable nearby AGN presented in \cite{AJ2025cl}: NGC\,1566, NGC\,2617, Mrk\,590, Mrk\,1018, and IRAS\,23226$-$3843. These objects were originally drawn from the larger sample presented in \cite{Temple2023} where they were identified as CSAGNs based on repeated optical spectroscopy from the BAT AGN Spectroscopic Survey (BASS)\footnote{\url{https://www.bass-survey.com/}}. Importantly, our analysis does not rely on the variable optical classifications of these sources, but it exploits the wide range of luminosities (and thus $\lambda_{\rm Edd}$) of these sources to explore the connection between spectral energy distribution and accretion rate.

We adopt the black hole mass ($M_{\rm BH}$) values primarily from the BASS DR2 catalog, which provides consistently derived estimates using uniform scaling relations and line-fitting procedures \citep[see][for details]{Koss2022,Mejia-Restrepo2022}. For NGC\,2617, we use the reverberation-based mass from \citet{Feng2021}. In BASS DR2, $M_{\rm BH}$ values for NGC\,1566, Mrk\,590, and Mrk\,1018 are based on single-epoch broad-line measurements, while that for IRAS\,23226--3843 is derived from the $M_{\rm BH}$–$\sigma_*$ relation \citep{koss2022b}.

In our previous studies (\citealp{AJ2026}, Jana et al. 2026b; hereafter Paper\,I and Paper\,II, respectively), we performed a detailed analysis of the UV/optical-to-X-ray spectral energy distributions (SEDs) of these five AGN. In the following, we provide a brief overview of our SED modeling procedure (see Paper\,II for details). We utilized a total of 1021 simultaneous multi-epoch, multi-wavelength observations from the \emph{Swift} observatory, using six UV/optical filters from UV optical telescope (UVOT; \citealp{Poole2008}), and X-ray telescope (XRT; \citealp{Burrows2005}) spectra covering the 0.5--10\,keV range. To improve the signal-to-noise ratio, we stacked several observations, resulting in a total of 214 observations (see Paper\,I for details). The UV/optical fluxes were corrected for host galaxy contamination using the host fluxes reported by \citet{Gupta2024}, who carried out a careful host/AGN decomposition using GALFIT \citep{Peng2010}.
The SED analysis was performed using a three-component phenomenological model, which included a \textsc{diskbb} component to represent thermal emission from the accretion disk \citep{Makishima1986}, a \textsc{blackbody} component for the soft X-ray excess, and a \textsc{cutoffpl} model for the primary Comptonized X-ray continuum, where we fixed the cutoff energy at 200\,keV, consistent with the median value of nearby AGN \citep{Ricci2018}. We also included a \textsc{pexrav} \citep{MZ98} component to reproduce reprocessed X-ray continuum emission and a Gaussian line at $\sim 6.4$\,keV to take into account the Fe K$\alpha$ emission feature. Dust extinction and line-of-sight absorption were incorporated using the \textsc{zdust} and \textsc{phabs} models, respectively. For all observations, with this model, we obtained a good fit (C-stat/degrees of freedom (dof) $\sim$ 1).
%A comprehensive description of the modeling and fitting procedures is provided in Paper\,II.

From the best-fit SED models, we calculated the accretion disk luminosity ($L_{\rm disk}$) by integrating the \textsc{diskbb} model over the 10$^{-7}$--0.5\,keV range (2.48\,nm -- 1.24\,cm), the soft-excess luminosity ($L_{\rm SE}$) by integrating the \textsc{blackbody} model over the 0.001--10\,keV range and the X-ray continuum luminosity ($L_{\rm PC}$) by integrating the \textsc{cutoffpl} component in the 0.1--500\,keV energy range. The total bolometric luminosity was then estimated as the sum: $L_{\rm bol} = L_{\rm disk} + L_{\rm SE} + L_{\rm PC}$, while the intrinsic X-ray continuum luminosity ($L_{\rm PC}^{\rm 2-10}$) was derived in the 2--10\,keV band. Finally, the X-ray bolometric correction was estimated as $\kappa_{\rm 2-10} = L_{\rm bol}/L_{\rm PC}^{\rm 2-10}$. The Eddington ratio is calculated as $\lambda_{\rm Edd}=L_{\rm bol}/L_{\rm Edd}$, where $L_{\rm Edd}=1.5\times10^{38}(M_{\rm BH}/M_{\odot})$ erg s$^{-1}$, which is appropriate for Solar metalicity gas. We refer the reader to Paper\,I \& II for detailed fitting procedure. We used \textsc{python} packages, \textsc{scipy} for statistical analysis and \textsc{linmix} \citep{kelly2009} for the regression analysis. We employed the Spearman correlation to test the significance of the correlation. For the statistical analysis, we used all available individual data points, while the binned points are shown only for visual guidance.
%is Eddington luminosity.

\begin{figure*}
\centering
\includegraphics[width=0.55\linewidth]{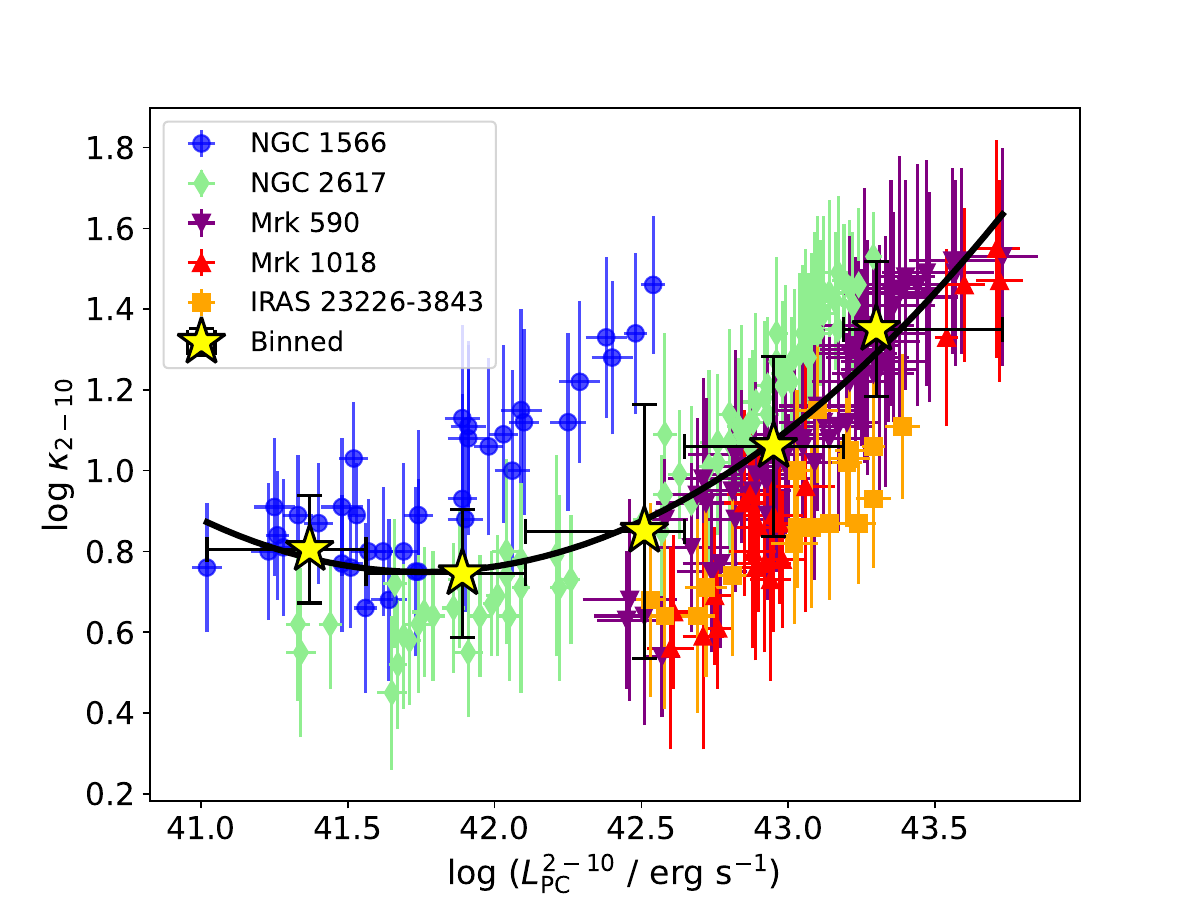}\hspace{-1cm}
\includegraphics[width=0.5\linewidth]{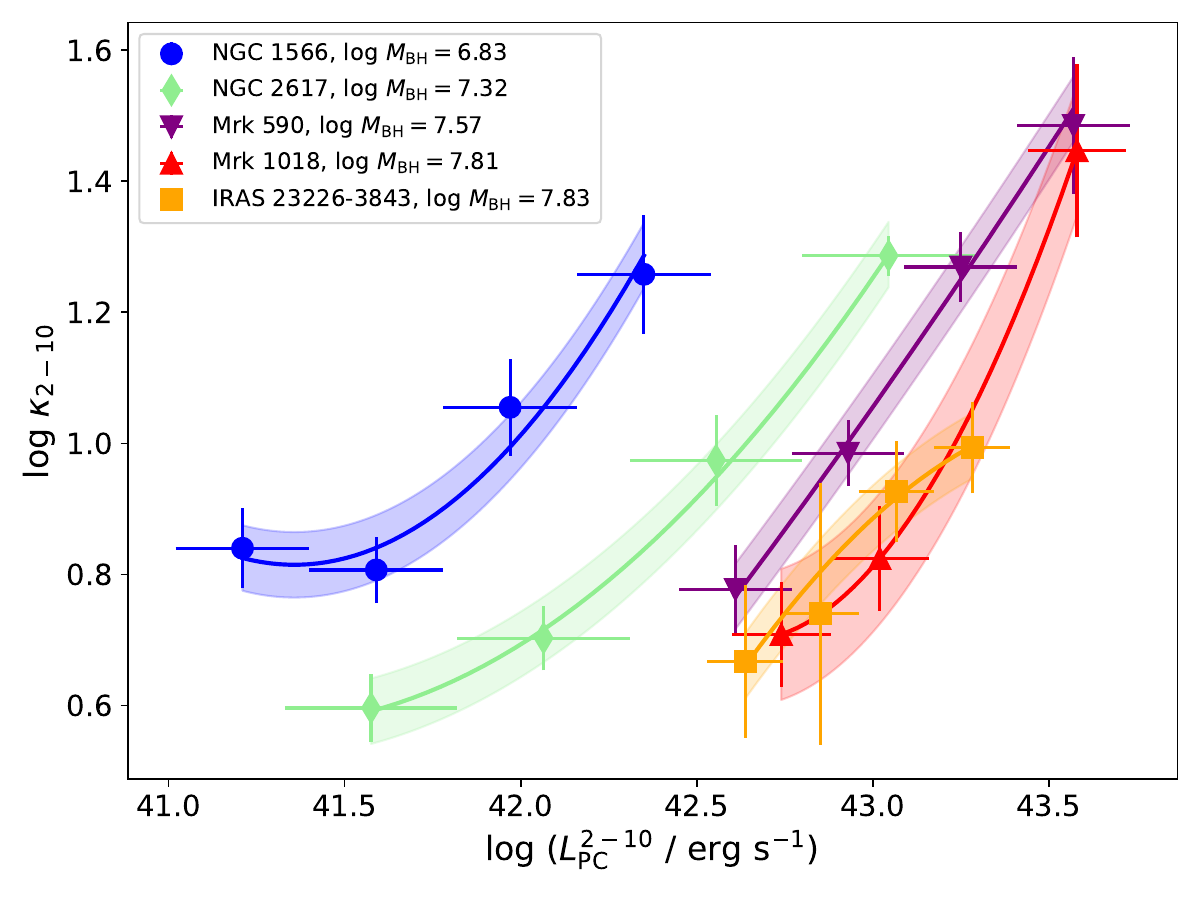}
\includegraphics[width=0.55\linewidth]{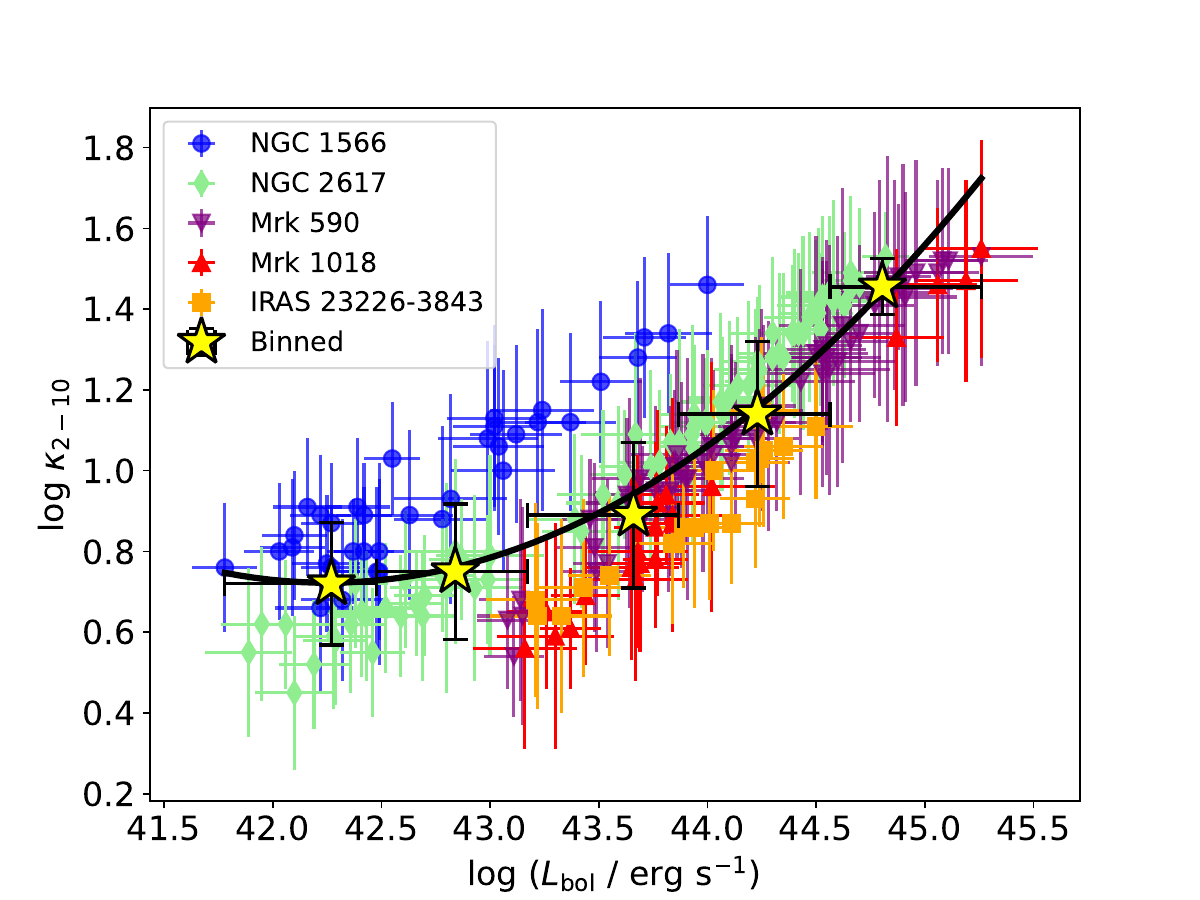}\hspace{-1cm}
\includegraphics[width=0.5\linewidth]{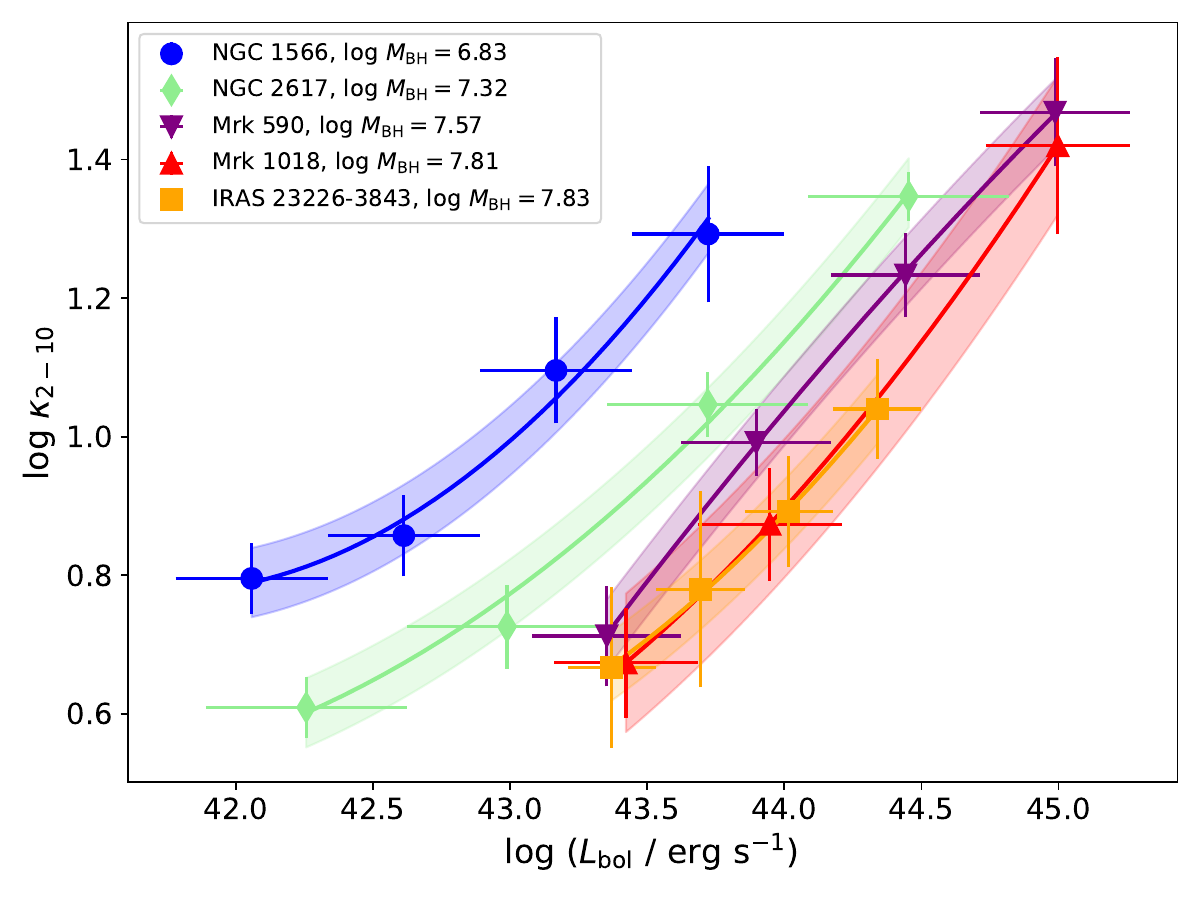}
\caption{Top left panel: Relation of X-ray bolometric correction ($\kappa_{\rm 2-10}$) with the 2--10\,keV X-ray continuum luminosity ($L_{\rm PC}^{\rm 2-10}$). The blue circles, green diamonds, purple down-triangles, red up-triangles, and orange squares represent NGC\,1566, NGC\,2617, Mrk\,590, Mrk\,1018, and IRAS\,23226--3843, respectively. The yellow stars represent the binned data point. The black line represents the best-fit (Eqn.~\ref{eqn:lx}) to the whole dataset. Top right panel: same as left panel, but with data re-binned for each source. The shaded regions represent the 1-$\sigma$ scatter. Bottom panels: same as top panels, but relation between $\kappa_{\rm 2-10}$ and $L_{\rm bol}$ (Eqn.~\ref{eqn:bol}).}
\label{fig:k-lum}
\end{figure*}

\section{The main parameter regulating X-ray Bolometric Corrections}
\label{sec:res}

In this section, we investigate the dependence of $\kappa_{\rm 2-10}$ on key physical parameters, such as $L_{\rm PC}^{\rm 2-10}$, $L_{\rm bol}$, $M_{\rm BH}$, and $\lambda_{\rm Edd}$. Across our sample, $\kappa_{\rm 2-10}$ spans a range of approximately $\sim 3-40$, while both $L_{\rm PC}^{\rm 2-10}$ and $L_{\rm bol}$ vary by $\sim 2.5$\,dex. Specifically, $L_{\rm PC}^{\rm 2-10}$ and $L_{\rm bol}$ range from $10^{41}$ to $10^{43.5}$ erg\,s$^{-1}$, and from $10^{41.5}$ to $10^{44}$ erg\,s$^{-1}$, respectively. The black hole mass covers one order of magnitude ($\log (M_{\rm BH}/M_{\odot})\sim 6.8-7.8$). The Eddington ratio spans about $\sim 3$\,dex with $\log \lambda_{\rm Edd}$ ranging between $-3.6$ and $-0.5$ for our sample.

\subsection{Dependence of $\kappa_{\rm 2-10}$ on luminosity and black hole mass}
\label{subsec:k-l-mbh}

The top left panel of Figure~\ref{fig:k-lum} presents the relation between $\kappa_{\rm 2-10}$ and $L_{\rm PC}^{\rm 2-10}$ for all sources. We find a strong positive correlation, with a Spearman rank coefficient of 0.78 and a p-value of $p \ll 10^{-10}$. The positive correlation is in agreement with previous studies \cite[e.g.,][]{Duras2020}. We first fitted the $\kappa_{\rm 2-10}-L^{\rm 2-10}_{\rm PC}$ relation using a linear regression model, with the fit returned with $\chi^2 = 212$ for 212 degrees of freedom (dof). However, this represents a simplification, as individual sources exhibit significant curvature and the ensemble shows substantial source to source dispersion. A second-order polynomial significantly improves the fit, 
with reducing $\chi^2$ by $\Delta \chi^2 = 65$ for one additional degree of freedom.
The $\kappa_{\rm 2-10}-L_{\rm PC}^{\rm 2-10}$ relation over the range $L_{\rm PC}^{\rm 2-10} = 10^{41-43.5}$ erg\,s$^{-1}$ is well described by a second-order polynomial:
\begin{multline}\label{eqn:lx}
\log \kappa_{\rm 2-10} = (0.781 \pm 0.021) + (0.131 \pm 0.042) \log L_{\rm PC,~42}^{\rm 2-10} \\
+(0.202 \pm 0.035) [\log L_{\rm PC,~42}^{\rm 2-10}]^2,
\end{multline}
where, $L_{\rm PC,~42}^{\rm 2-10}=L_{\rm PC}^{\rm 2-10}/10^{42}~{\rm erg\,s^{-1}}$.
The right panel of Figure~\ref{fig:k-lum} shows the bolometric corrections of each AGN rebinned for different luminosity ranges, with each source independently fit with the same quadratic function. Shaded regions represent the 1$\sigma$ intrinsic scatter for each fit. Each bin contains the data points lying within the interval of $L_{\rm PC}^{\rm 2-10}$, and we use their median value in the $\kappa_{\rm 2-10}$ to represent the bin. We find that, while all the sources can be fitted with a second-order polynomial function, the slope and curvature terms are broadly similar across the sample, while the zero-point offset differs from source to source, and it systematically increases with $M_{\rm BH}$. 

Bottom panel of Figure~\ref{fig:k-lum} shows the $\kappa_{\rm 2-10}$–$L_{\rm bol}$ relation. $L_{\rm bol}$ shows strong positive correlation with $\kappa_{\rm 2-10}$ ($p \ll 10^{-10}$) for individual sources, as well as for the whole sample. Several previous studies have also reported a positive correlation between $\kappa_{\rm 2-10}$ and $L_{\rm bol}$ \citep[e.g.,][]{Vasudevan2007,Lusso2012,Duras2020,Gupta2025}. Similar to what we found for the $L_{\rm PC}^{\rm 2-10}$, the $\kappa_{\rm 2-10}-L_{\rm bol}$ relation in the range of $L_{\rm bol}=10^{41.5-44}$ \eps is best described by a second-order polynomial fit:
\begin{multline}\label{eqn:bol}
\log \kappa_{\rm 2-10} = (0.729 \pm 0.032) - (0.058 \pm 0.026) \log (L_{\rm bol,~42}) \\
+(0.111 \pm 0.024) [\log (L_{\rm bol,~42})]^2.
\end{multline}
Here $L_{\rm bol,~42}=(L_{\rm bol})/10^{42}~{\rm erg~s^{-1}}$.
The linear fit yields $\chi^2 = 195$ for 212 degrees of freedom, while the second-order polynomial model provides a substantially improved description of the data, reducing the fit statistic by $\Delta \chi^2 = 69$ for one additional dof. The right bottom panel of Fig.~\ref{fig:k-lum} presents the binned data, with individual quadratic fits per source. As in the $L_{\rm PC}^{\rm 2-10}$ case, the slopes are consistent however a zero-point offset in $L_{\rm bol}$ is observed across different sources. With the offset in $L_{\rm bol}-\kappa_{\rm 2-10}$ increases with $M_{\rm BH}$. The relations suggest that for a given $L_{\rm bol}$, $\kappa_{\rm 2-10}$ decreases with increasing $M_{\rm BH}$.

We find no evidence for a direct correlation between $\kappa_{\rm 2-10}$ and $M_{\rm BH}$ across the sample. Our results are in agreement with the recent work of \cite{Gupta2025}, who also found no relation between $\kappa_{\rm 2-10}$ and $M_{\rm BH}$ for a wide range of $M_{\rm BH}$ ($\sim 10^{5.5-10}~M_{\odot}$).

\subsection{The strong dependence on the Eddington ratio}
\label{subsec:k-edd}
The dependence of the X-ray bolometric corrections on both luminosity and black hole mass, as shown in Figure~\ref{fig:k-lum}, suggests that the Eddington ratio might play a fundamental role. The $\kappa_{\rm 2-10}$ and $\lambda_{\rm Edd}$ are found to be tightly correlated with Spearman correlation of 0.97 with $p \ll 10^{-10}$. As both $\kappa_{\rm 2-10}$ and $\lambda_{\rm Edd}$ explicitly depend on $L_{\rm bol}$, an apparent correlation between them may arise partly from this dependence. To assess their intrinsic connections, we performed a partial correlation analysis among $\kappa_{\rm 2-10}$, $L_{\rm bol}$, and $\lambda_{\rm Edd}$. We obtained the following partial correlation coefficients:
\\
\\
$L_{\rm bol}-\kappa_{\rm 2-10}: -0.56$ (controlling for $\lambda_{\rm Edd}$).\\
$L_{\rm bol}-\lambda_{\rm Edd}: 0.83$ (controlling for $\kappa_{\rm 2-10}$).\\
$\lambda_{\rm Edd}-\kappa_{\rm 2-10}: 0.93$ (controlling for $L_{\rm bol}$).\\

This shows that, even after accounting for the coupling introduced by $L_{\rm bol}$, $\lambda_{\rm Edd}$ and $\kappa_{\rm 2-10}$ remain strongly correlated, implying that the Eddington ratio is the dominant physical driver linking the $\kappa_{\rm 2-10}$ and $L_{\rm bol}$.

\begin{figure*}
\centering
\includegraphics[width=0.55\linewidth]{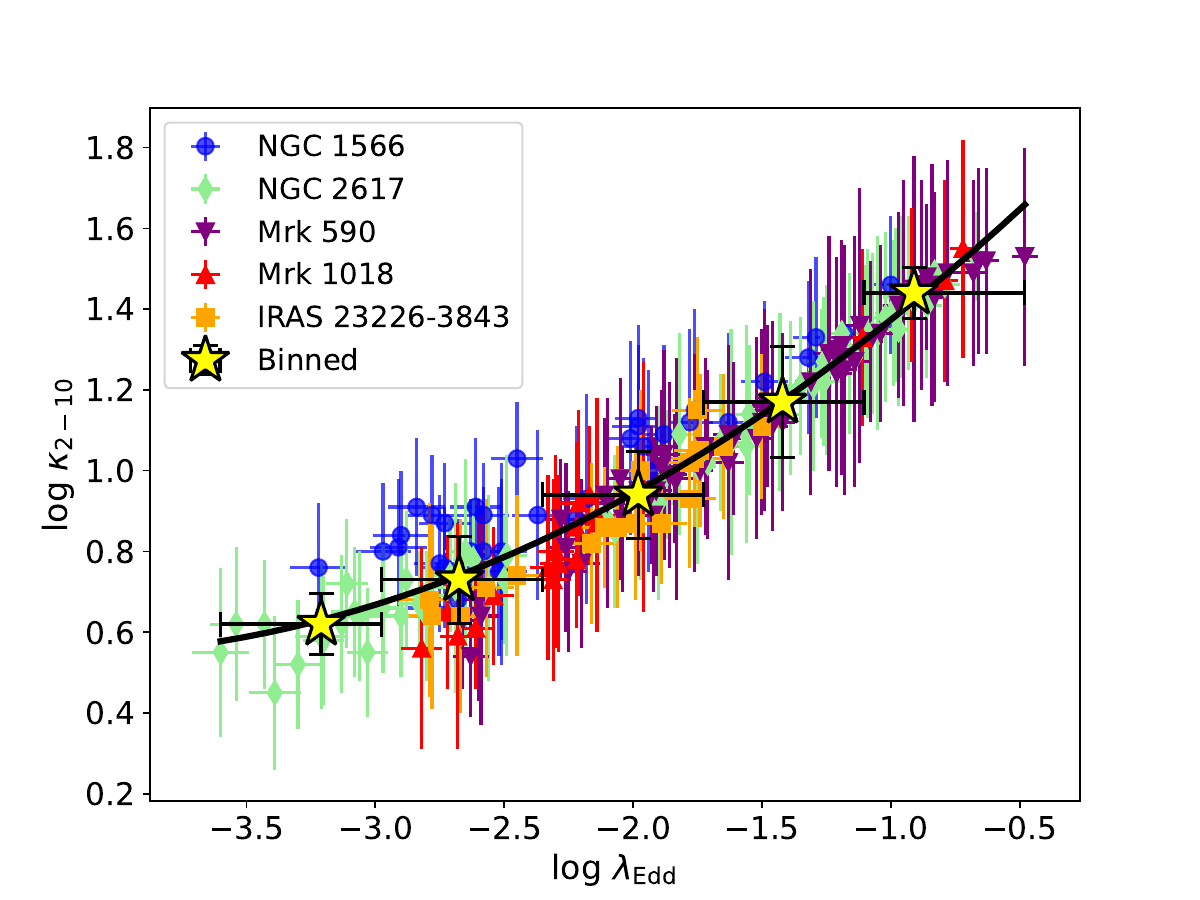}\hspace{-1cm}
\includegraphics[width=0.5\linewidth]{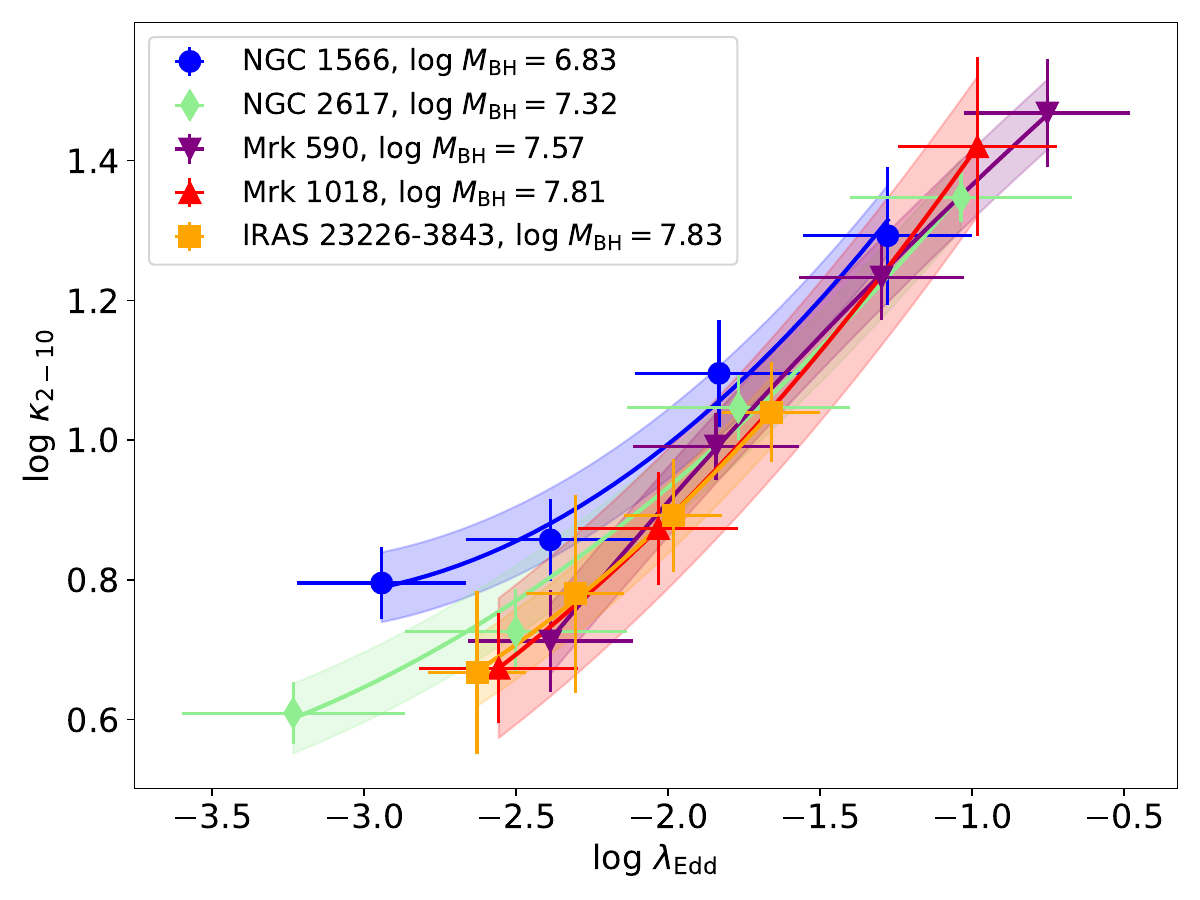}
\caption{Top panel: Variation of X-ray bolometric correction ($\kappa_{\rm 2-10}$) as a function of Eddington ratio ($\lambda_{\rm Edd}$). The blue circles, green diamonds, purple down-triangles, red up-triangles, and orange squares represent the data points from NGC\,1566, NGC\,2617, Mrk\,590, Mrk\,1018, and IRAS\,23226--3843, respectively. The yellow stars represent the binned data point. The black line represents the best-fit (Eqn.~\ref{eqn:edd}) to the data. Bottom panel: same as top panel, but with median value in bins of $\log \lambda_{\rm Edd}$. The shaded regions represent the 1-$\sigma$ scatter.}
\label{fig:k-ed}
\end{figure*}

Figure~\ref{fig:k-ed} displays the relation between $\kappa_{\rm 2-10}$ and $\lambda_{\rm Edd}$. A linear regression analysis returned with a good fit with $\chi^2= 94$ for 212 dof, for $\lambda_{\rm Edd}-\kappa_{\rm 2-10}$ relation. The linear regression analysis yields $\log \kappa_{\rm 2-10}=(1.69 \pm 0.08) + (0.36 \pm 0.04) \log \lambda_{\rm Edd}$. A second-order polynomial fit improves the fitting by $\Delta \chi^2 = 28$ for one additional dof in the range $\lambda_{\rm Edd}\sim 10^{-3.6}-10^{-0.5}$, yielding the following best-fit relation:

\begin{multline}\label{eqn:edd}
\log \kappa_{\rm 2-10} = (1.957 \pm 0.058) + (0.654 \pm 0.072) \log \lambda_{\rm Edd} \\
+ (0.074 \pm 0.022) (\log \lambda_{\rm Edd})^2.
\end{multline}

The bottom panel of Figure~\ref{fig:k-ed} displays binned data and individual quadratic fits for each source. Although the fit parameters obtained by fitting individual sources exhibit some variation, they remain consistent within uncertainties, confirming a strong and very tight correlation between $\kappa_{\rm 2-10}$ and $\lambda_{\rm Edd}$ across the sample.

Previous studies also found positive correlations, albeit with large scatter \citep[$\sim 0.3-0.5$\,dex;][]{Vasudevan2007,Lusso2012,Duras2020,Gupta2025}. However, focusing on a small number of CSAGNs, which cover three orders of magnitude in $\lambda_{\rm Edd}$, significantly reduces the scatter and reveals a much tighter correlation (1-$\sigma$ scatter of 0.05\,dex) than previously reported, while also extending the relation to lower $\lambda_{\rm Edd}$. The tight $\kappa_{\rm 2-10}-\lambda_{\rm Edd}$ relation indicates that that the relative strength of the X-ray emission depends strongly on the Eddington ratio.

When the X-ray continuum is excluded from the bolometric luminosity (i.e. by considering only the disk and soft-excess emission), the intrinsic scatter in the $\kappa_{\rm 2-10}-\lambda_{\rm Edd}$ relation increases from 0.05\,dex to 0.11\,dex. This indicates that some covariance contributes to the small scatter in the original $\kappa_{\rm 2-10}$--$\lambda_{\rm Edd}$ relation, although the correlation remains intrinsically tight and therefore still likely reflects a physical trend. This interpretation is further supported by the changing fractional contributions of the disk, warm corona, and hot corona across accretion rate (see Appendix~\ref{sec:fraction}).

\subsection{The relation between X-ray bolometric correction factor and $\alpha_{\rm OX}$}
\label{subsec:aox_kx}

\begin{figure*}
\centering
\includegraphics[width=0.55\linewidth]{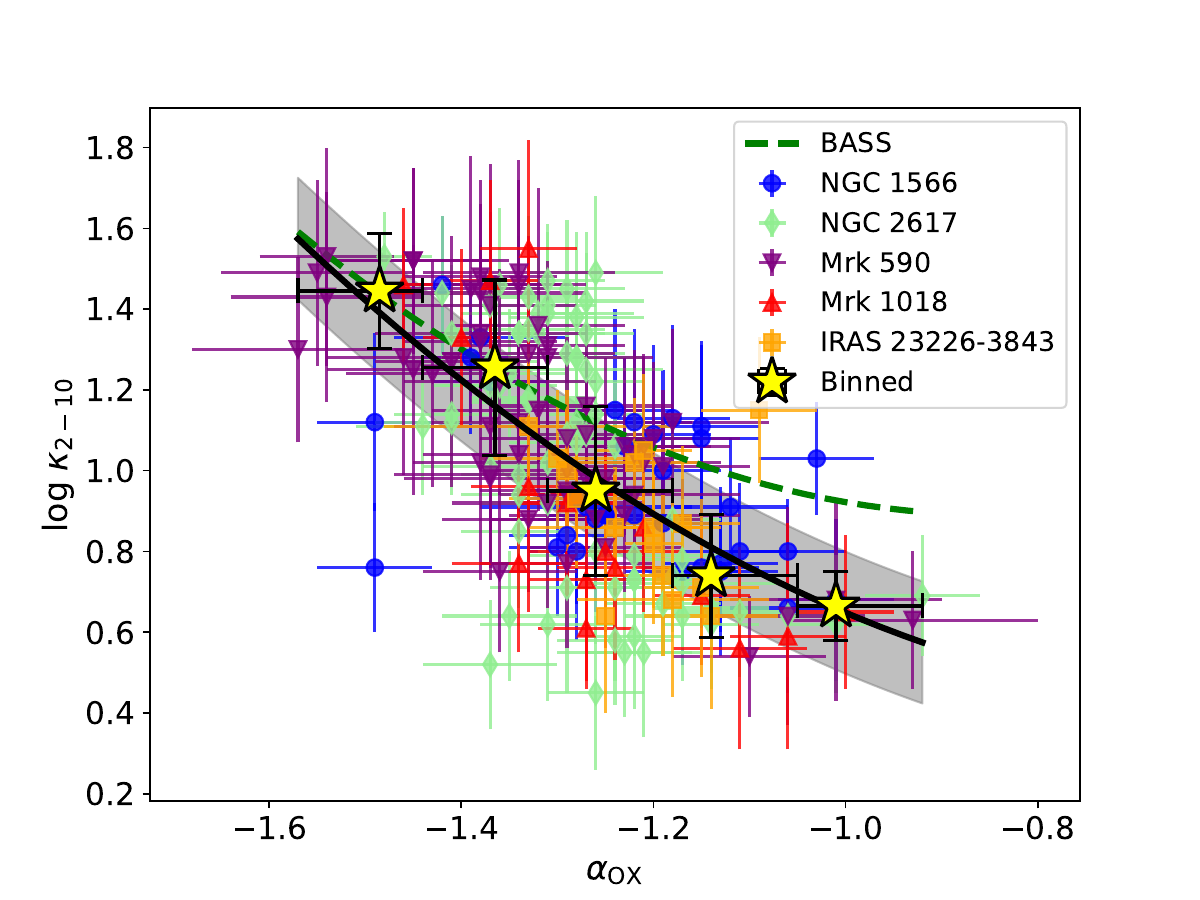}\hspace{-1cm}
\includegraphics[width=0.5\linewidth]{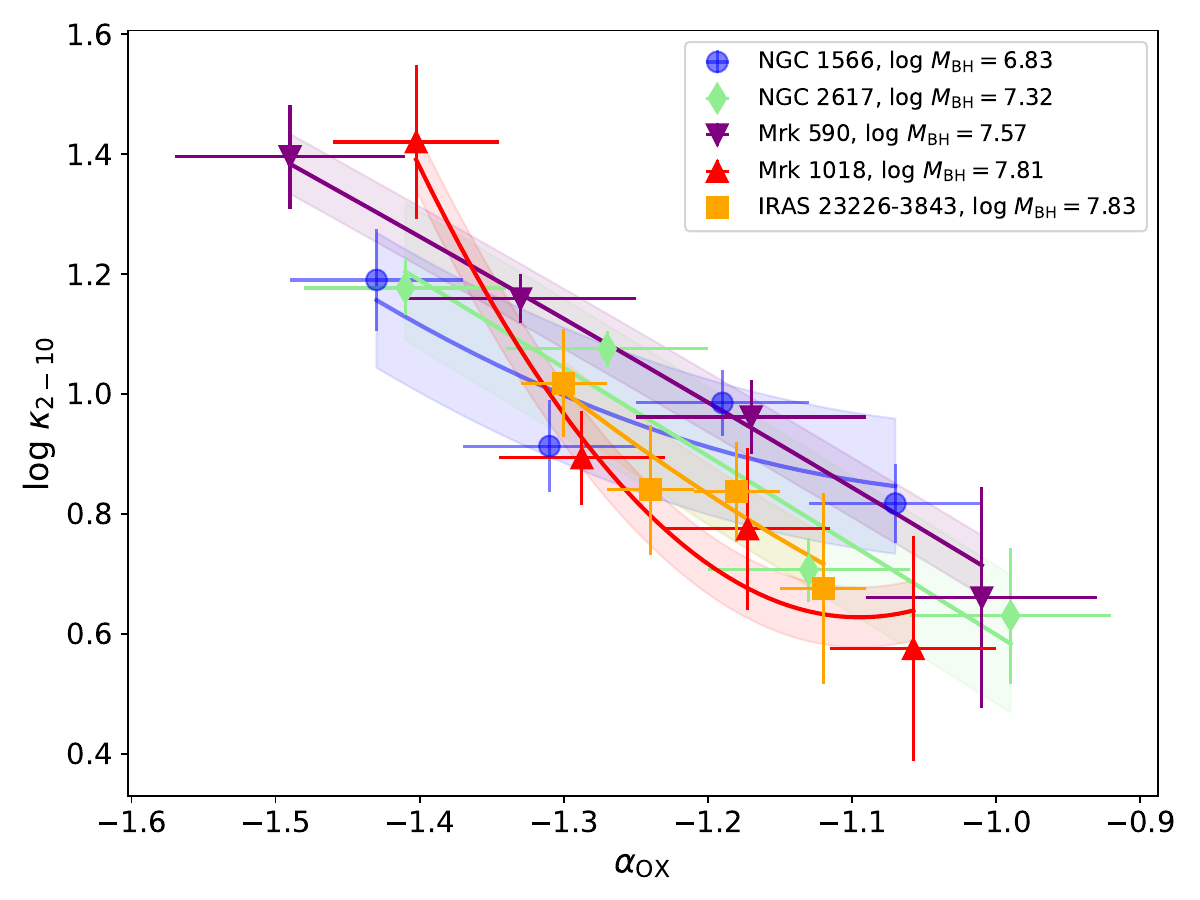}
\caption{Left panel: Variation of X-ray bolometric correction ($\kappa_{\rm 2-10}$) as a function of UV-to-X-ray spectral index ($\alpha_{\rm OX}$). The blue circles, green diamonds, purple down-triangles, red up-triangles, and orange squares represent the data points from NGC\,1566, NGC\,2617, Mrk\,590, Mrk\,1018, and IRAS\,23226--3843, respectively. The yellow stars represent the binned data point. The black line represents the best-fits (Eqn.~\ref{eqn:alpha}) of the data. 
The green dashed line represents the $\kappa_{\rm 2-10}-\alpha_{\rm OX}$ relation for unobscured BASS AGNs \citep{Gupta2024}.
Right panel: same as right panel, but with binned data. The shaded regions represent the 1-$\sigma$ scatter.}
\label{fig:k-aox}
\end{figure*}

The UV-to-X-ray spectral index ($\alpha_{\rm OX}$) is widely used as diagnostic of accretion disk-corona coupling, providing useful information on the relation between the accretion disk and the corona using only two monochromatic fluxes \citep[e.g.,][]{Shemmer2006,Just2007}. To explore this connection with our dataset, we studied the relation between $\alpha_{\rm OX}$ and the $\kappa_{\rm 2-10}$, shown in Figure~\ref{fig:k-aox}. We find a negative correlation between these quantities with a Spearman coefficient of $-0.66$ with $p \ll 10^{-10}$, consistent with previous studies \citep[e.g.,][]{Lusso2010,Gupta2024}. A linear regression analysis yields a good fit with $\chi^2 = 195$ for 212 dof to the $\kappa_{\rm 2-10}-\alpha_{\rm OX}$ relation, however, individual sources show a curve, suggesting a polynomial fit. A second-order polynomial fit improved the fit $\Delta \chi^2 = 5$ for 1 dof (2.2$\sigma$) over the linear fit. The 
$\kappa_{\rm 2-10}$–$\alpha_{\rm OX}$ relation is described by a quadratic function,
\begin{multline}\label{eqn:alpha}
\log \kappa_{\rm 2-10} = (0.162 \pm 0.062) + (0.314 \pm 0.091) \alpha_{\rm OX} \\
+(0.769 \pm 0.231) \alpha_{\rm OX}^2.
\end{multline}
with an intrinsic scatter of 0.18\,dex, in the range $\alpha_{\rm OX} \sim -0.9$ to $–1.5$. 
The right panel of Fig.~\ref{fig:k-aox} displays $\kappa_{\rm 2-10}$ for each AGN, with each source fit independently with a second-order polynomial. While individual AGNs are fitted with slightly different coefficients, their $\kappa_{\rm 2-10}$–$\alpha_{\rm OX}$ relations are consistent within the uncertainties.

Our result is broadly consistent with the study of \citet{Gupta2024} and \cite{Lusso2010} for $\alpha_{\rm OX} > -1.2$, but shows significant deviations at higher $\alpha_{\rm OX}$. Specifically, we find a monotonic decrease in $\kappa_{\rm 2-10}$ for all the values of $\alpha_{\rm OX}$, while \cite{Gupta2024} shows a steady value of $\kappa_{\rm 2-10}$ at $\alpha_{\rm OX}>-1.2$.

\section{Discussion}
\label{sec:disc}

The X-ray bolometric correction serves as a critical parameter to constrain the balance between the high-energy emission from the X-ray corona and the optical/UV radiation produced by the accretion disk. In this study, we investigate the dependence of $\kappa_{\rm 2-10}$ on several fundamental physical parameters, namely $L_{\rm PC}^{\rm 2-10}$, $L_{\rm bol}$, $M_{\rm BH}$, and $\lambda_{\rm Edd}$. Our analysis is based on a sample of five extremely variable AGN, observed over 1,000 epochs with simultaneous optical, UV, and X-ray coverage.

We find that, for each source, the X-ray bolometric correction positively correlates with both $L_{\rm PC}^{\rm 2-10}$ and $L_{\rm bol}$ (Fig~\ref{fig:k-lum}). However, the relations between $\kappa_{\rm 2-10}$ and luminosities for the ensemble of all sources exhibit significantly larger scatter compared to the correlation with $\lambda_{\rm Edd}$, due to substantial offsets between sources. Specifically, the $L_{\rm bol}-\kappa_{\rm 2-10}$ relation shows a typical scatter of $\sim0.11-0.18$\,dex for individual sources compared to $\sim0.31$\,dex for the whole sample. The individual tracks systematically shift toward higher luminosities with increasing black hole mass (see right panels of Figure~\ref{fig:k-lum}). In contrast, the $\kappa_{\rm 2-10}-\lambda_{\rm Edd}$ relations exhibit a much tighter trend both for individual sources and the entire sample. This finding implies that $\lambda_{\rm Edd}$ is the fundamental parameter governing the fraction of X-ray emission in AGNs. The partial correlation analysis among $L_{\rm bol}$, $\kappa_{\rm 2-10}$ and $\lambda_{\rm Edd}$ also suggests this, showing the strongest $\kappa_{\rm 2-10}-\lambda_{\rm Edd}$ correlation, after controlling for $L_{\rm bol}$. A consistent result was recently obtained by \citet{Gupta2025}, who showed that between $L_{\rm bol}$ and $\lambda_{\rm Edd}$, the latter is the primary regulator of $\kappa_{\rm 2-10}$, based on the disappearance of the luminosity-dependence of $\kappa_{\rm 2-10}$ at low $\lambda_{\rm Edd}$.

\begin{figure*}
\centering
\includegraphics[width=0.75\linewidth]{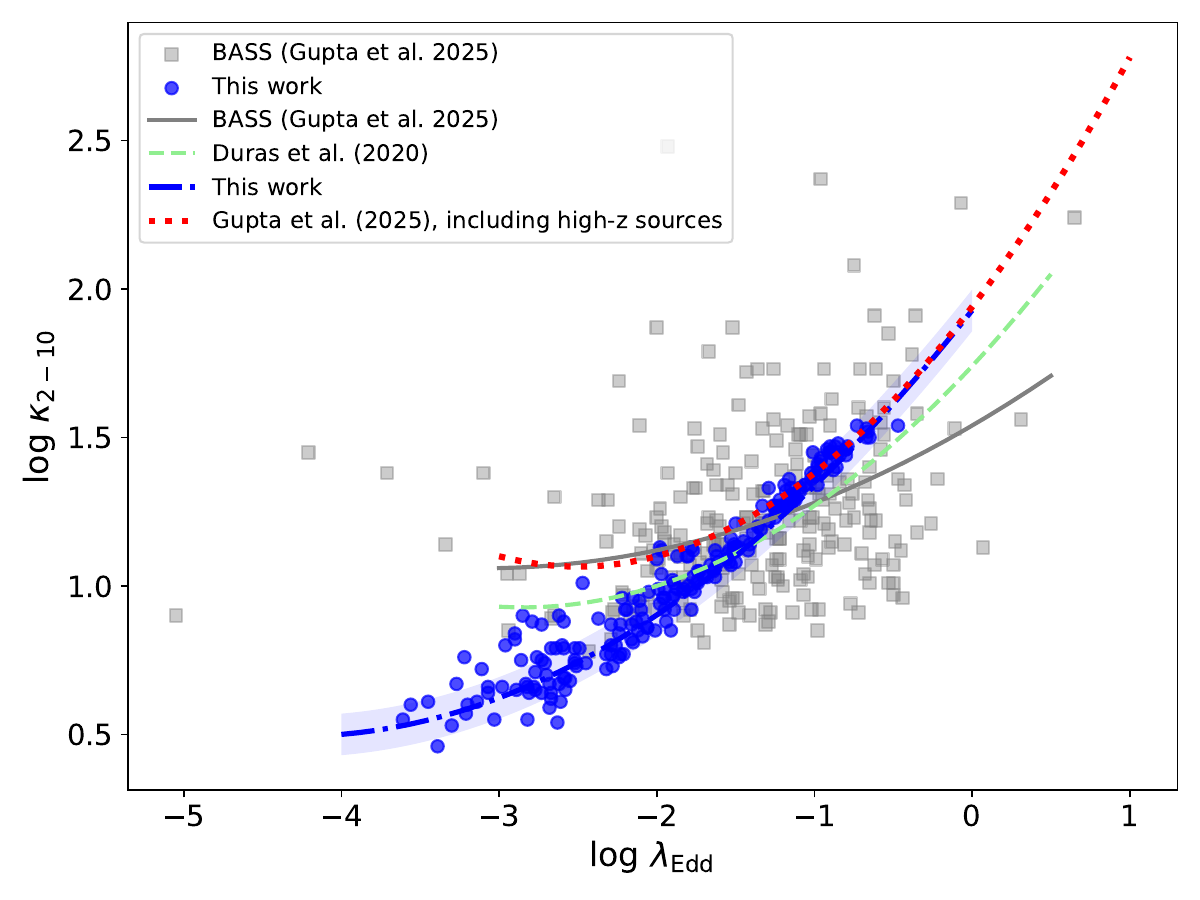}
\caption{Blue circles and gray squares represent the variation of the X-ray bolometric correction ($\kappa_{\rm 2-10}$) as a function of the Eddington ratio ($\lambda_{\rm Edd}$) for the variable AGN studied here, and for unobscured AGNs in BASS from \citet{Gupta2025}, respectively. The blue dashed-dot and orange solid lines represent the best-fit for our sample (Eqn.\,\ref{eqn:edd}) and unobscured BAT AGNs (Eqn.\,2 of \citealp{Gupta2025}),
respectively. The red dotted line shows the best fitted relation of $\kappa_{\rm 2-10}-\lambda_{\rm Edd}$ for unobscured BAT AGNs with high redshift quasars, taken from \citet{Gupta2025}. The green dashed line shows the $\kappa_{\rm 2-10}-\lambda_{\rm Edd}$ relation from \cite{Duras2020}. The blue shaded-region represent the 1-$\sigma$ scatter from the present study.}
\label{fig:all}
\end{figure*}

Figure~\ref{fig:all} displays the variation of $\kappa_{\rm 2-10}$ with $\lambda_{\rm Edd}$, where we compare our results with those obtained by previous studies. The blue circles represent the observations from the current work, while the gray squares are the unobscured Swift/BAT AGNs from \cite{Gupta2025}. Our results match closely with \cite{Duras2020} and with the work of \citeauthor{Gupta2025} (\citeyear{Gupta2025}; where high-z quasars are included) for $\lambda_{\rm Edd} \sim 0.01-0.1$ and $\lambda_{\rm Edd}>0.01$, respectively. We notice that the works of both \cite{Gupta2025} and \cite{Duras2020} show a flattening at $\lambda_{\rm Edd}<0.01$, with their $\kappa_{\rm 2-10}$ being significantly larger than what we found here. On the other hand, focusing on low-luminosity AGNs ($L_{\rm bol}<10^{42}$ \eps), \citet{Lopez2024} found a X-ray bolometric corrections around $\kappa_{\rm 2-10} \simeq 9.6\pm 1.0$, which deviates from our findings.

Earlier works have reported a positive correlation between $\kappa_{\rm 2-10}$ and $\lambda_{\rm Edd}$, but with considerably larger intrinsic scatter of 0.25\,dex to 0.5\,dex \citep{Vasudevan2009,Lusso2012,Duras2020}. 
The scatter was generally attributed to a combination of intrinsic AGN variability, heterogeneity in source properties, and the use of non-simultaneous multi-wavelength data. Even with simultaneous optical/UV-to-X-ray SEDs for nearby AGN, \cite{Gupta2025} found a scatter of $\sim 0.3$\,dex for the $\kappa_{\rm 2-10}-\lambda_{\rm Edd}$ relation. The authors argued that the scatter could be due to the BH spin or inclination angle which were not constrained in the SED fitting. In contrast to previous studies, we find a remarkably small intrinsic scatter of 0.05\,dex in the $\kappa_{\rm 2-10}-\lambda_{\rm Edd}$ relation. This is achieved by concentrating on five well-studied CSAGNs with extensive simultaneous multi-epoch, multi-wavelength coverage. By limiting the sample size, we are likely minimizing systematic uncertainties and suppressing the impact of additional parameters such as black hole spin or inclination. We also note that the $M_{\rm BH}$ in our sample span a relatively narrow range (6.8–7.8\,dex ), which could reduce the scatter compared to population studies, where the broader mass distribution and the larger associated uncertainties in $M_{\rm BH}$ contribute significantly to the observed scatter. The exceptionally small scatter may also suggest that the sources in our sample have similar disk inclinations, resulting in comparable apparent radiative output and viewing geometries because of the anisotropic nature of disk emission. Since Type\,1 AGNs typically have inclination angles $\leq 45^{\circ}$ \citep[e.g.,][]{Netzer2015}, the associated geometric effect could introduce a scatter of up to $\sim 0.15$\,dex for a large sample. For a smaller sample with more uniform inclinations, this scatter would naturally be reduced.

We also investigated if $\alpha_{\rm OX}$ traces $L_{\rm bol}$ in AGNs. A positive correlation between $\alpha_{\rm OX}$ and $\kappa_{\rm 2-10}$ is found, which is expected as $\alpha_{\rm OX}$ correlates with both UV and X-ray luminosities \citep[e.g.,][]{Steffen2006,Lusso2010,Marchese2012,Stalin2010,Grupe2010}.
We note that $\alpha_{\rm OX}$ can serve as a reliable proxy for estimating $L_{\rm bol}$ without requiring full multi-wavelength coverage. The relatively small scatter (0.18\,dex) in the $\alpha_{\rm OX}–\kappa_{\rm 2-10}$ relation suggests that Eqn.\,~\ref{eqn:alpha} provides an effective tool to trace $L_{\rm bol}$.
However, we caution that the $\alpha_{\rm OX}-\lambda_{\rm Edd}$ relation might show a `$\Lambda$’-shaped trend (Fig.\,5 of paper\,II, see also \citealp{Ruan2019}), implying that a given value of $\alpha_{\rm OX}$ may correspond to two possible $\kappa_{\rm 2-10}$ values, which could introduce ambiguity in estimating $L_{\rm bol}$ for sources near the turnover region. Due to this fact, the $\kappa_{\rm 2-10}-\alpha_{\rm OX}$ relation shows larger scatter than $\kappa_{\rm 2-10}-\lambda_{\rm Edd}$.

\subsection{Implication on the accretion process in SMBHs}
\label{subsec:implication}

Accretion theory suggests that for $\lambda_{\rm Edd} \sim 0.01$–$0.3$, the standard geometrically-thin, optically-thick disk model \citep{SS73} may provide an adequate description of black holes accreting at moderate rates, where the flow remains radiatively efficient and the thin-disk approximation holds. While this model may serve as a useful first approximation for SMBHs, it fails to reproduce several fundamental signatures of AGN emission. For example, it predicts blue optical/UV spectral slopes and prominent Lyman edges that are not observed \citep[e.g.,][]{Koratkar1999}; the strong variability seen in CSAGNs on short timescales is inconsistent with the much longer viscous timescales expected for a thin disk (e.g., \citealp{Stern2018,Ricci2023Nat}); and microlensing studies reveal surface brightness profiles far dimmer than those predicted for an optically thick flow \citep{Morgan2010,Chelouche2019}. At lower ($\lambda_{\rm Edd} \lesssim 0.01$) and higher ($\lambda_{\rm Edd} \gtrsim 0.3$) Eddington ratios, it is instead expected that the flow deviates strongly from the thin-disk solution and becomes radiatively inefficient \citep[e.g.,][]{Abramowicz1988,Ho2008,Ho2009,Jin2012a,Lusso2012,Yuan2014,Li2024b}.

Observational evidence gathered over the past decades has shown that the properties of the innermost regions of AGN, traced by X-ray emission, are closely linked to the Eddington ratio. In particular, the photon index ($\Gamma$) of the primary power-law decreases with increasing $\lambda_{\rm Edd}$ up to $\sim 0.01$, and then shows a positive correlation with $\lambda_{\rm Edd}$ at higher Eddington ratios (e.g., \citealp{Shemmer2006,Gou2009,Diaz2023}). This trend has been clearly confirmed by our recent study focused on the same extremely variable AGNs studied here \citep{AJ2026}. The soft excess also exhibits a dependence on Eddington ratio. While survey-based studies have suggested a possible increase of the soft excess intensity relative to the continuum emission with $\lambda_{\rm Edd}$ for $\lambda_{\rm Edd} \gtrsim 0.01$  \citep{Boissay2016,Gliozzi2020}, dedicated investigations of changing-state AGNs have revealed a strong weakening, and in some cases the disappearance of the soft excess at $\lambda_{\rm Edd} \lesssim 0.01$ \citep{Noda2018,Ghosh2022,Mehdipour2022}. This behavior was recently confirmed through the detection of a very clear linear correlation between the ratio of soft-excess to power-law flux and the Eddington ratio, extending down to $\lambda_{\rm Edd} \simeq 0.001$ \citep{AJ2026}. Below this threshold, the soft excess systematically disappears, based on the same set of observations analyzed here \citep[see also,][]{Hagen2024,Kang2025}. 
Recently, \citet{Middei2026} showed that the UV and X-ray emissions show a tight correlation and follow a non-linear relation above $\lambda_{\rm Edd} \gtrsim 0.01$, while the relation breaks below it.
All of this could be related to a change in the structure of the accretion flow at $\lambda_{\rm Edd} \sim 0.01$, with the accretion disk transitioning into an advection-dominated accretion flow \citep[e.g.,][]{Yuan2014}. Recent studies carried out using physical SED models confirmed that at low Eddington ratios ($\lambda_{\rm Edd}\lesssim0.01$) AGNs exhibit dramatic spectral changes, consistent with a collapse of the inner disk into a hot, inefficient plasma \citep[e.g.,][]{Jin2012a,Younes2019,Hagen2024,Kang2025}. Such a change was also clearly observed in the SEDs of the CSAGNs studied here (Paper\,II).

The tight positive correlation of $\kappa_{\rm 2-10}-\lambda_{\rm Edd}$ (see Fig.~\ref{fig:k-ed}) shows that the relative contribution of X-ray emission to $L_{\rm bol}$ decreases with increasing $\lambda_{\rm Edd}$. The fraction of X-ray emission (in 2--10\,keV band) is observed to increase from $\sim 3\%$ at high $\lambda_{\rm Edd}$ to $\sim 30\%$ at low $\lambda_{\rm Edd}$. This trend could be associated to a weaker X-ray corona, possibly due to a more efficient cooling at high $\lambda_{\rm Edd}$. Such a trend is closely related to the so-called X-ray weakness phenomenon observed in high accreting AGNs and quasars \citep[e.g.,][]{Dong2012,Inayoshi2025,Tortosa2026}. This result is particularly relevant for JWST-selected high-$z$ sources, which may preferentially lie at high $\lambda_{\rm Edd}$. The combination of steeper spectra and larger $\kappa_{\rm 2-10}$ at high $\lambda_{\rm Edd}$ implies a lower expected 2–10\,keV flux at fixed UV/optical emission. Therefore, at least part of the reported X-ray weakness in the JWST high-$z$ population may reflect accretion-state effects rather than requiring extreme obscuration (e.g., \citealp{Maiolino2025}).

At low Eddington ratio ($\lambda_{\rm Edd}<0.01$) the $\kappa_{\rm 2-10}$ declines with decreasing $\lambda_{\rm Edd}$, with the X-rays becoming stronger with respect to the optical/UV emission. This is consistent with previous studies wherein low-luminosity AGNs are observed to be `X-ray loud' compared to their UV emission \citep[e.g.,][]{Ho1999,Ho2008}. This phenomena could be associated with a change in the accretion flow, with the X-ray emission being dominated by emission from a hot, radiatively inefficient flow or extended corona, while the disk is truncated at a large distance from the SMBH. At this $\lambda_{\rm Edd}$, the seed photons could be supplied by the jet base, or corona itself via synchrotron emission \citep{Markoff2001,Yuan2014}. Interestingly, by fitting the $\kappa_{\rm 2-10}-\lambda_{\rm Edd}$ trend with two linear relations we find a break at $\log \lambda_{\rm Edd} = -2.36 \pm 0.15$, consistent with the break observed in the trend between $\Gamma$ and $\lambda_{\rm Edd}$ \citep{AJ2026}.

These systematic dependencies point to a fundamental restructuring of the accretion flow with $\lambda_{\rm Edd}$, and call for a unified physical framework to interpret them. In the flux-frozen disk scenario, \citet{Hopkins2025} predicts that components such as the BLR, torus, soft X-ray excess, and corona emerge self-consistently and evolve with $\lambda_{\rm Edd}$. The tight $\kappa_{\rm 2-10}$–$\lambda_{\rm Edd}$ correlation we find is consistent with this general picture, reinforcing the view that $\lambda_{\rm Edd}$ is the key parameter governing the disk–corona balance and driving structural transitions in the flow.

\section{Summary}
\label{sec:summary}

We studied optical/UV-to-X-ray SEDs of five unobscured CSAGNs, using more than 1000 simultaneous observations, obtained with \emph{Swift}/UVOT and \emph{Swift}/XRT. To improve the signal-to-noise ratio, we stacked adjacent observations, resulting in a total of 214 observations (Paper\,II). We investigate how the X-ray bolometric correction, $\kappa_{\rm 2-10}$ changes with several key AGN parameters, namely $M_{\rm BH}$, $L_{\rm PC}^{\rm 2-10}$, $L_{\rm bol}$ and $\lambda_{\rm Edd}$. 
The strong variability of these CSAGNs enables us to obtain robust constraints on the bolometric corrections even at low Eddington ratios ($\log \lambda_{\rm Edd} < -3$), a regime that has not been systematically explored before. In the following, we summarize our findings.

\begin{enumerate}
\item We find a tight positive correlation of $\kappa_{\rm 2-10}$ with $L_{\rm PC}^{\rm 2-10}$, $L_{\rm bol}$ and $\lambda_{\rm Edd}$ (see Fig.~\ref{fig:k-lum} and ~\ref{fig:k-ed}; Eqns.\,\ref{eqn:lx}--\ref{eqn:edd}), consistent with previous studies \citep[e.g.,][]{Duras2020,Gupta2025}.

\item Individual sources follow distinct tracks in the $\kappa_{\rm 2-10}$ versus $L_{\rm PC}^{\rm 2-10}$ and $L_{\rm bol}$ planes, with these tracks systematically shifting toward higher luminosities for larger black hole masses (Fig.~\ref{fig:k-lum}).

\item When considering the relationship between $\kappa_{\rm 2-10}$ and $\lambda_{\rm Edd}$ we find that both individual sources and the entire sample follow the same remarkably tight correlation (Eqn.\,\ref{eqn:edd}), with a much smaller scatter (0.05\,dex; Fig.~\ref{fig:k-ed}) than previous studies. This provides clear evidence that $\lambda_{\rm Edd}$ is the fundamental parameter regulating the fraction of X-ray emission in AGNs, and suggests a very tight relation between $\kappa_{\rm 2-10}$ and $\lambda_{\rm Edd}$, consistent with the previous results of \cite{Gupta2025}.
The strong non-linear positive $\kappa_{\rm 2-10}-\lambda_{\rm Edd}$ relation implies that UV/optical emission accounts for over 90\% of the bolometric output at high Eddington ratios, while X-rays contribute up to 30\% at low $\lambda_{\rm Edd}$.

\item $\alpha_{\rm OX}$ correlates tightly with $\kappa_{\rm 2-10}$ ($\sigma \sim $ 0.18\,dex, Fig.\,\ref{fig:k-aox} and Eqn.\,\ref{eqn:alpha}), showing that $\alpha_{\rm OX}$ can be reliably used as a proxy to estimate $\kappa_{\rm 2-10}$ and $L_{\rm bol}$ in AGNs.

\item The observed dependence of the bolometric correction on $\lambda_{\rm Edd}$, together with the changes in the X-ray spectral shape and the soft excess intensity, indicates a clear change in the accretion flow with $\lambda_{\rm Edd}$. Changes in the X-ray spectral shape (i.e., photon index and soft excess intensity, see \citealp{AJ2026}, A. Jana et al. 2026b) show that this restructuring becomes evident around $\lambda_{\rm Edd} \sim 0.01$, consistent with the transition from a standard thin disk to a radiatively inefficient hot flow observed in black hole X-ray binaries \citep[e.g.,][]{RM06,Done2007}.

\end{enumerate}

Our results highlight the importance of coordinated time-domain, multi-wavelength observations of extremely variable AGNs. These objects allow for direct tracking of changes in the accretion flow structure, as reflected in both X-ray and optical/UV fluxes and spectral properties (e.g., \citealp{Temple2023,Li2024b,Panda2024}, Jana et al. 2025b,c). 

The tight correlation observed between $\kappa_{\rm 2-10}$ and $\lambda_{\rm Edd}$ provides a reliable empirical framework for estimating bolometric luminosities in AGNs, especially in cases where full SED coverage is unavailable. In a forthcoming publication, we will investigate super-Eddington AGNs ($\lambda_{\rm Edd} \geq 1$) to extend our analysis of the accretion properties of nearby AGNs to this regime. Several current and upcoming facilities, such as LSST \citep{Ivezic2019LSST} and \emph{NewAthena} \citep{Cruise2025newAthena} will be crucial for multi-epoch, multi-wavelength observations aimed at probing the accretion physics of SMBHs in detail, particularly at high redshift.

%%%%%%%%%%%%%%%%%%%%%%%%%

\begin{acknowledgments}
We thank the anonymous referee for providing insightful comments and suggestions that improved this manuscript.
We acknowledge the usage of data and/or software provided by the High Energy Astrophysics Science Archive Research Center (HEASARC), which is a service of the Astrophysics Science Division at NASA/GSFC and the High Energy Astrophysics Division of the Smithsonian Astrophysical Observatory.
AJ acknowledge support from ANID grants FONDECYT Postdoctoral fellowship 3230303.
CR acknowledges support from SNSF Consolidator grant F01$-$13252, Fondecyt Regular grant 1230345, ANID BASAL project FB210003 and the China-Chile joint research fund. 
KKG acknowledges financial support from the Belgian Federal Science Policy Office (BELSPO) in the framework of the PRODEX Programme of the European Space Agency.
BT acknowledges support from the European Research Council (ERC) under the European Union's Horizon 2020 research and innovation program (grant agreement number 950533).
AT acknowledges support from the INAF Large Program ``DELUX" of the ``Ricerca Fondamentale 2024" INAF program.
LCH was supported by the National Science Foundation of China (12233001) and the China Manned Space Program (CMS-CSST-2025-A09).
FEB acknowledges support from ANID-Chile BASAL CATA FB210003 and FONDECYT Regular 1241005.
YD acknowledge support from ANID grants FONDECYT Postdoctoral fellowship 3230310.
AT acknowledges support from the INAF Large Program “DELUX” of the “Ricerca Fondamentale 2024” INAF program.
\end{acknowledgments}

\begin{contribution}
%%This section gives authors the space to recognize author contributions. The text inside this environment is NOT counted towards the total word quanta. At a minimum, manuscripts are expected to include this text:

All authors contributed to the content of the publication.
AJ and CR wrote the first draft of the paper. FEB, KKG, BT provided scientific comments, suggestions and edited the draft at various stages. AT, RL, RM, HKC, YD, GD, KK, MK, SP, and MT provided helpful scientific comments and suggestions. AT and GD provided data for the work.
%% But authors are expected to provide more specific details, e.g. 
%%
%%SC was responsible for writing and submitting the manuscript.
%%WWM came up with the initial research concept and edited the manuscript.
%%OTS obtained the funding and edited the manuscript.
%%EBF provided the formal analysis and validation. He also edited the manuscript.
%%GEH Supervised the undergraduates, wrote the software and administers the project github and Zenodo repositories.
%%
%% Authors can use the Contributor Role Taxonomy (CRediT) at
%% for ideas on how write a good statement tailored to their needs.
%FEB provided scientific comments, suggestions and edits to the manuscript at various stages. 

\end{contribution}

\facilities{Swift(XRT and UVOT).}
\software{SciPy \citep{Vincentelli2020}, Linmix \citep{kelly2009}.} 

%\software{astropy \citep{2013A&A...558A..33A,2018AJ....156..123A,2022ApJ...935..167A},  
%          Cloudy \citep{2013RMxAA..49..137F}, 
%          Source Extractor \citep{1996A&AS..117..393B}
%          }


\appendix

\section{Fraction of Emission from Different Components}
\label{sec:fraction}
\begin{figure}
\centering
\includegraphics[width=0.5\linewidth]{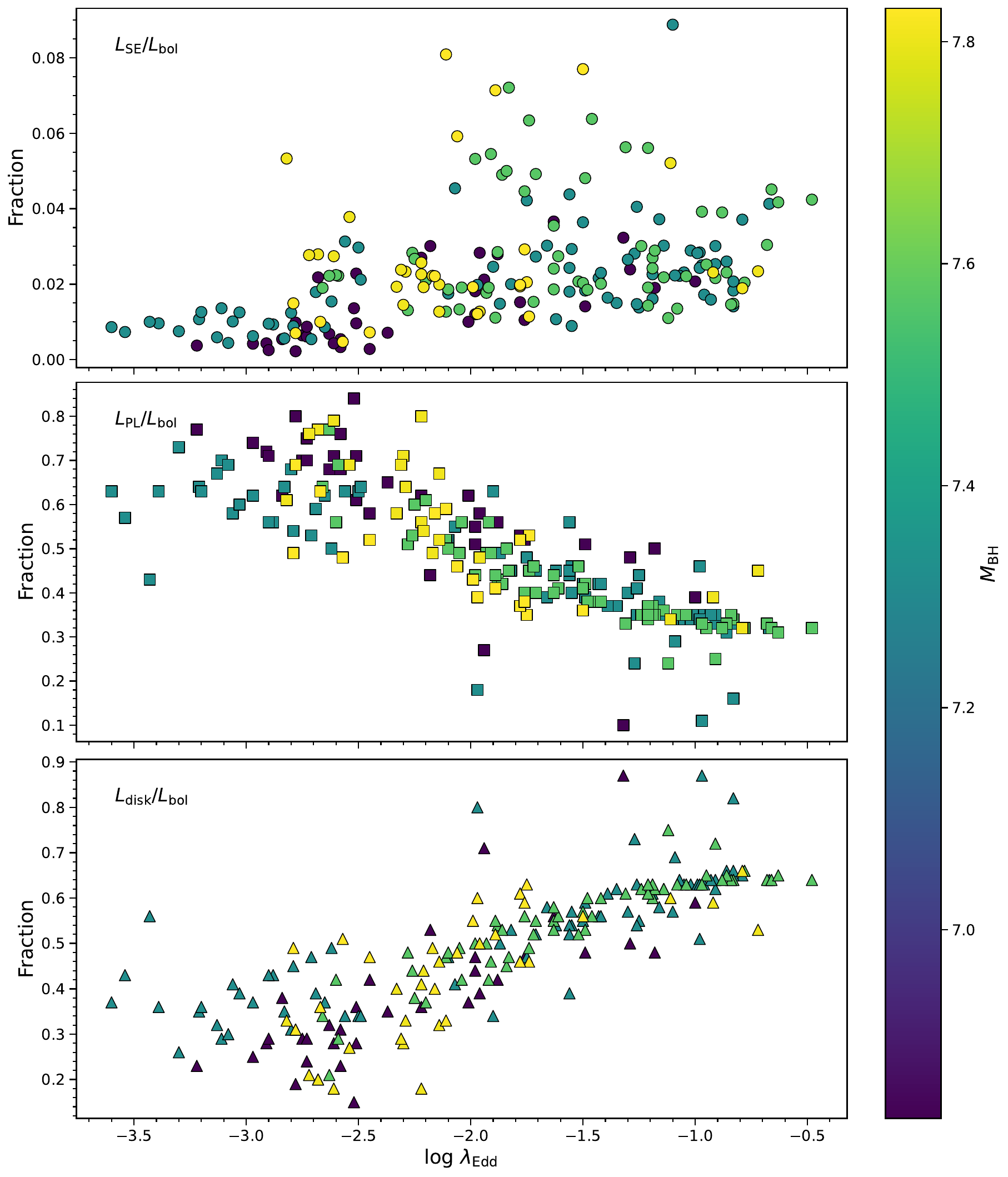}
\caption{raction of total emission from i) warm corona, ii) hot corona, and iii) accretion disk as a function of Eddington ratio is shown in top, middle and bottom panel, respectively. The emission from the warm corona, hot corona, and accretion is estimated in $10^{-7}-0.5$\,keV, $0.001-10$\,keV, and $0.1-500$\,keV, respectively.}
\label{fig:fraction}
\end{figure}

Figure~\ref{fig:fraction} presents the fractional contributions of the warm corona, hot corona, and accretion disk to the total bolometric luminosity as a function of Eddington ratio. In the low-accretion regime ($\lambda_{\rm Edd}<0.01$), the hot corona contributes up to $\sim 50-80$\% of the total radiative output, whereas in the high-accretion regime its contribution decreases to $\sim 20$\%. The warm corona contributes up to $\sim 10$\% of the bolometric luminosity, but its contribution declines markedly toward low Eddington ratios. In contrast, at high accretion rates ($\lambda_{\rm Edd}>0.01$), the accretion disk contributes up to $\sim 60-70$\% of the total radiative power.

These trends demonstrate that the dominant contributor to the bolometric luminosity changes systematically with accretion rate. The hot corona dominates the radiative output in the low-accretion state, whereas the accretion disk becomes the principal contributor in the high-accretion state. We therefore interpret the tight $\kappa_{\rm 2-10}-\lambda_{\rm Edd}$ relation as reflecting a genuine change in the broadband energy budget across accretion states, rather than arising solely from mathematical coupling.

\bibliography{ref-clagn}{}
\bibliographystyle{aasjournalv7}



\end{document}